\documentclass[12pt]{iopart}
\usepackage{graphicx}
\usepackage{iopams}
\begin{document}

\title{Gelation as arrested phase separation in short-ranged attractive colloid-polymer mixtures}

\author{Emanuela Zaccarelli$^{1,2}$, 
Peter J. Lu$^3$, Fabio Ciulla$^1$, David. A. Weitz$^{3}$ and Francesco Sciortino$^{1,2}$}
\address{ $^1$ Dipartimento di Fisica and $^2$CNR-INFM-SOFT, 
Universit\`{a} di
Roma La Sapienza, P.le A. Moro 2, I-00185 Rome, Italy\\
$^3$Department of Physics, and SEAS, Harvard
University, Cambridge, MA 02138, USA
} 
\eads{\mailto{emanuela.zaccarelli@phys.uniroma1.it}, \mailto{plu@fas.harvard.edu}}

\begin{abstract}
We present further evidence that gelation is an arrested
phase separation in attractive colloid-polymer mixtures, based
on a method combining confocal microscopy experiments with
numerical simulations recently established in {\bf Nature 453,
499 (2008)}. Our results are independent of the form of the
interparticle attractive potential, and therefore should apply
broadly to any attractive particle system with short-ranged,
isotropic attractions. We also give additional characterization
of the gel states in terms of their structure, inhomogeneous
character and local density.
\end{abstract}

\section{Introduction}
The addition of non-adsorbing polymers to a colloidal suspension can
induce an effective depletion attraction between the colloidal
particles\cite{Asa58a}, leading to a much richer phase
behavior\cite{Ile95aPRE, Lek92a} than that normally observed for
atomic or molecular systems \cite{And02a}.  Colloid-polymer mixtures
are therefore often used as model systems of
complex fluids, which are normally encountered in technological
applications such as food and personal care products, where the
attractive interactions may significantly impact product stability.
Micron-sized constituent colloids are large enough to be imaged with light,
while still small enough to have their dynamics driven by $k_B
T$, the thermal energy, where $T$ is the temperature and $k_B$,
the Boltzmann constant. Both the range and magnitude of the
attractive interactions can by finely tuned by adjusting the
radius $r_p$ and concentration $c_p$ of the added polymers.
With sufficiently high $c_p$, these mixtures form gels,
i.e.  networks of particles that span across the volume in a
disordered arrested structure, which can sustain shear stresses,
despite the small packing fraction $\phi$ occupied by the colloidal
particles\cite{Tra04a,Cip05a,Zac07a}.
In the limit of strong
attractions and very small $\phi$, gelation is driven by the
diffusion-limited cluster aggregation (DLCA) of fractal clusters of
colloidal particles, a mechanism well-studied in the
past\cite{Vicsekbook,Wei85a,Car92a}. In the limit where $\phi$ is
large, particles arrest to form colloidal glasses\cite{Sci05a}, which can be
of two fundamental types: repulsive, driven only by packing and
geometrical frustration\cite{Pus87a}, and attractive, when particle
bonds become dominant, at sufficiently high attraction
strength\cite{Daw00a,Pha02a}. A simple extension of attractive glasses
to lower and lower densities is not necessarily a good explanation for
gel formation, due to the intervening of liquid-gas (colloid
rich-colloid poor) phase separation.  Several different mechanisms have been proposed for the gel
transition at low- and intermediate-$\phi$ and high interaction
strengths\cite{Zac07a}. These include
percolation\cite{Grant93,Sha03a}, mode-coupling theory (MCT) attractive
glass\cite{Ber03a}, cluster-MCT\cite{Cat04aJPCM}, fluid-crystal
transition\cite{Pus93a}, spinodal-induced
gelation\cite{Asn97,Pou99a,Sci93a}, viscoelastic phase
separation\cite{Tan00aJPCM}.  Stimulated by a series of numerical 
studies\cite{Zaccapri,Fof05b,Cha07a} and by analogies with studies of
atomic fluids\cite{Sas00PRL} and polymer solutions\cite{Leo87a}, the hypothesis of  arrested spinodal
decomposition has recently  received a particular attention in experimental 
studies\cite{Man05a,Buz07a,Car07a}.  One of the greatest
challenges in reaching an unambiguous interpretation is to find
a way to precisely quantify the attractive interaction potential, which
would allow a quantitative connection to be forged between
observed experimental phenomena and a well-characterized
thermodynamic phase or state.

Thus far,  the mapping of the experimental $c_p$ into a
system-independent thermodynamic parameter has been based on
assumptions of a specific shape for the interparticle
 potential $V(r)$,
such as the Asakura-Oosawa (AO) model\cite{Asa58a} typically employed to
describe colloid-polymer mixtures.  Within this model,
colloids (with radius $a$) behave as hard spheres, while
polymers forming coils in solution have a dual nature: with
respect other polymers, they are an ideal gas in which polymer
coils can interpenetrate; with respect to the colloids,
polymers behave as hard spheres of radius $r_p$, which cannot
interpenetrate the colloidal spheres. With these simple
assumptions, an effective colloid-colloid attractive potential
is derived,
\begin{equation}
V_{AO}(r)= k_B T \frac{\partial \Pi}{\partial V}
F(r/2(a+r_p))
\label{eq:AO}
\end{equation} 
where $\Pi$ is the osmotic pressure, $V$ the volume and $F(x)$ a well
defined function\cite{Asa58a} which vanishes for $x>1$. The maximum strength of
attraction occurs when two colloidal spheres touch, and this attraction is denoted by the parameter 
$U \equiv V_{AO}(r=a) /k_B T$.  To facilitate comparison between different experimental
conditions, the concentration-dependent radius of the polymer is
typically divided by the colloid radius to yield a dimensionless size
ratio $\xi \equiv r_p / a$.  A universal gelation state diagram may
then be parameterized by $\phi$, $\xi$ and $U$, yielding a way to
compare different experiments and theories\cite{Lu06a}. However, while
$\phi$ and $\xi$ are unambiguously defined in experiment, simulation
and theory, $U/k_B T$ depends entirely on the shape of the interaction
potential. This model-dependence of $U$ leaves its quantification
vulnerable to the uncertainties in the applicability of a given
theory. 

Recently Noro and Frenkel\cite{Nor00aJCP} have established a
generalized law of corresponding states for systems interacting with
short-range attractive potentials ($\xi \lesssim 0.1$), in terms only of the normalized
second virial coefficient $B_2^* \equiv (3/8a^3) \int_0^\infty
(1-e^{-V(r)/k_B T}) r^2 dr$. For a given value of $B_2^*$, any system
interacting with any given short-ranged attractive potential (e.g. AO,
square well or generalized Lennard-Jones) shares the same
thermodynamic properties and structure. Hence, either a direct measurement of
$B_2^*$ or a connection of the experimental attraction strength (e.g. expressing $c_p$ in terms of $B_2^*$) would allow a universal
description of the system. For example, one
recent work on lysozyme has used the Noro-Frenkel law to
express the crystallization transition under various solvent
conditions in terms of a universal $B_2^*$\cite{Sed07a}. 
Another possibility often exploited in the past is to rely on
 the well-known Baxter model\cite{Bax68a}, which
corresponds to the sticky hard sphere limit: the attraction
range becomes infinitesimal while the attraction strength becomes infinite in
such a way that the second virial coefficient $B_2$ remains
finite. This model is theoretically convenient because it is amenable to analytic treatment within
Percus-Yevick (PY) closure\cite{hansen06} and, consequently, it has
been used for a long time in comparison to experiments\cite{Che94a}. 
However, the
thermodynamic inconsistency of the PY solution has often led to an
inaccurate location of the gel line  with respect to the phase-separation line in the thermodynamic phase diagram
\cite{Ver94a}. Only a recent accurate location of the gas-liquid phase
separation, which was obtained numerically by Miller and
Frenkel\cite{Mil03a}, has opened the way for a more careful use of the Baxter
model, via the $B_2^*$ scaling, in relation to experiments
 where samples were shown to reach equilibrium over the course of several months\cite{Buz07a}.

Only a truly model-independent measure of the attractive
potential, that could be applied to a variety of attractive
colloidal systems, would facilitate quantitative comparison
between theory and experiments, and help to elucidate the
mechanisms driving the onset of gelation. To address this issue,
in a recent paper\cite{Lu08a}, we have introduced a 
novel method for generic short-range isotropic attractive 
interactions --- independent of any specific potential shape or 
model --- that is capable of quantitatively mapping the 
experimental observations of gel formation in colloid-polymer mixtures onto the corresponding
thermodynamic phase diagram.  Through the comparison of the cluster
mass distributions, measured directly at the single-particle level using confocal microscopy and in
simulations of various short-ranged attractive potentials, we have
established a one-to-one relationship between the experimental $c_p$
and $B_2^*$, building on the Noro and Frenkel law\cite{Nor00aJCP}. In doing so, we
provide unambiguous evidence that gelation takes
place precisely where spinodal decomposition occurs, evidencing a
strong link between thermodynamic behaviour and dynamic arrest. The
gel is triggered from the initial separation into two phases, of which
the densest may become locally arrested due to an intervening attractive
glass transition\cite{Zac07a}, and hence kinetics only intervenes
after thermodynamics has provided the necessary density fluctuations.

In the present manuscript, we report additional analysis of the
experimental system in both the fluid and the gel phase to complement the
results previously reported in\cite{Lu08a}. These results provide further
evidence that gelation in short-ranged attractive
colloid-polymer mixtures is an arrested phase separation.

\section{Description of the experimental system}
Sterically-stabilized colloidal spheres of polymethylmethacrylate
(PMMA) with DiIC$_{18}$ fluorescent dye are suspended in a 5:1 (by
mass) solvent mixture of bromocyclohexane (CXB, Aldrich) and
decahydronaphthalene (DHN, Aldrich) for several months, following 
procedures already described in \cite{Lu06a,Lu08a,Lu07a}.  The PMMA particles have
radius $a=560\pm10$ nm, determined with dynamic light
scattering. Tetrabutylammonium chloride (TBAC, Fluke) is added until
saturation ($\sim$4 mM) to screen any residual long-range Coulombic
repulsion. The colloidal suspension is carefully
buoyancy-matched\cite{Lu08a}.  Two sets of depletant polystyrene (PS)
are used, with molecular weights respectively of $M_W$=69.2 kDa and
$M_W$=681 kDa. From Zimm plots of static light scattering data, the
radii of gyration of the two polymers are determined to be 10.0
and 33.0 nm respectively, yielding $\xi=0.018$ and $\xi=0.059$ in
the two cases, always well within the short-range attraction
limit where the Noro-Frenkel law is valid.  The overlap
concentrations, defined as $c_p^* \equiv 3 M_W / 4 \pi r_p^3 N_A$, are
27.2 and 7.5 mg/mL, respectively for the two sets of polymers, with
$N_A$ being the Avogadro's number.  In all cases, the raw polymer
concentrations are directly measured as the mass ratio of mg PS per g
of total sample mass, then converted into a $\phi$-dependent
free-volume $c_p$ (mg/ml)\cite{Lek92a}.  Samples at varying polymer
concentration $c_p$ are generated while preserving constant $\phi$,
although a small decrease is observed with increasing polymer
concentration\cite{Lu08a};  this may be due to larger clusters sticking to the walls of the 
macroscopic sample chamber.

Following our previously-reported imaging
protocol\cite{Lu06a,Lu08a,Lu07a}, 3D stacks of 181 images are
collected, each an 8-bit TIFF of 1000 x 1000 pixels, at 10 frames per
second.  Each image stack covers a volume of $60 \times 60 \times 60$
$\mu$m$^3$, taken from the center of the sample at least $20\mu$m away
from any sample chamber surface to minimize edge effects. 
Each fluid sample is homogenized and
equilibrated for four hours, then 26 independent 3D image
stacks are collected with an automated confocal
microscope\cite{Lu07a}. Each gel sample is homogenized and then
observed immediately, with 3D stacks of the same sample volume
collected every 50 seconds for the first 5000 seconds, then
every 1000 seconds for the subsequent 100,000 seconds.
\begin{figure}[tbh]
\begin{center}
\includegraphics[width=9cm,angle=0.,clip]{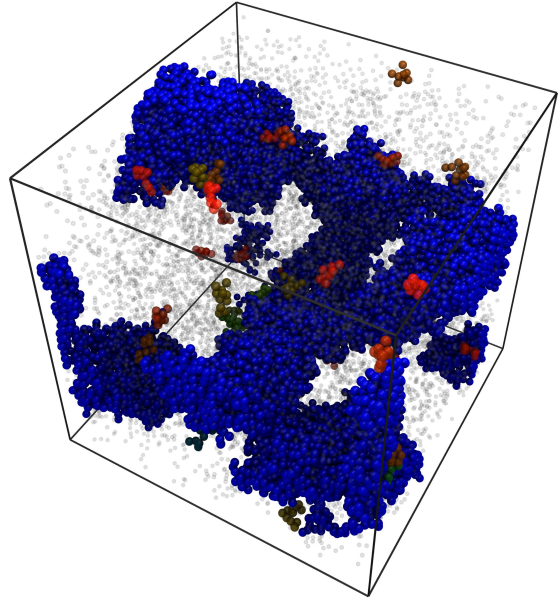}
\end{center}
\caption{3d-Reconstruction of the gel for $\phi=0.14$, $\xi=0.059$
  and $c_p=1.80$ mg/ml. Isolated clusters are shown in different colors for
  different cluster sizes, while monomers are depicted as transparent
  (and smaller) spheres. }
\label{fig:gel}
\end{figure}

Using confocal microscopy, we observe and individually locate
each colloidal particle in the sample imaging volume in 3D; a
typical gel is shown in Fig.~\ref{fig:gel}. Most
of the colloidal particles are connected in a spanning network,
while at all times a residual fraction of particles is present in
monomers and smaller clusters that freely diffuse throughout the voids of the
network. 
We also observe a continuous exchange between these unconnected 
particles and the gel network, due to the finite attraction strength of order a few $k_B T$ . 
However, once a percolating gel network forms (which happens generally within a few 
thousand seconds), it persists for all experimentally observed time --- it is arrested. 

Locating each particle individually with confocal microscopy
also allows more detailed structural analysis, such as a direct
calculation of the static structure factor, defined as
$S(q)\equiv \langle|\sum_{i=1}^{N} e^{i \bf{q} \cdot
  \bf{r}_j}|^2\rangle / N$.  For fluid samples, we average over the
$26$ independent configurations. For gel samples, we follow a single
configuration over time. We calculate $S(q)$ for a cubic box of
size $50 \mu$m, centered within the sample imaging volume, in order to
minimize edge effects, which if present,
will only affect the range $2qa \lesssim 0.2$.

\begin{figure}[tbh]
\begin{center}
\includegraphics[width=7cm,angle=0.,clip]{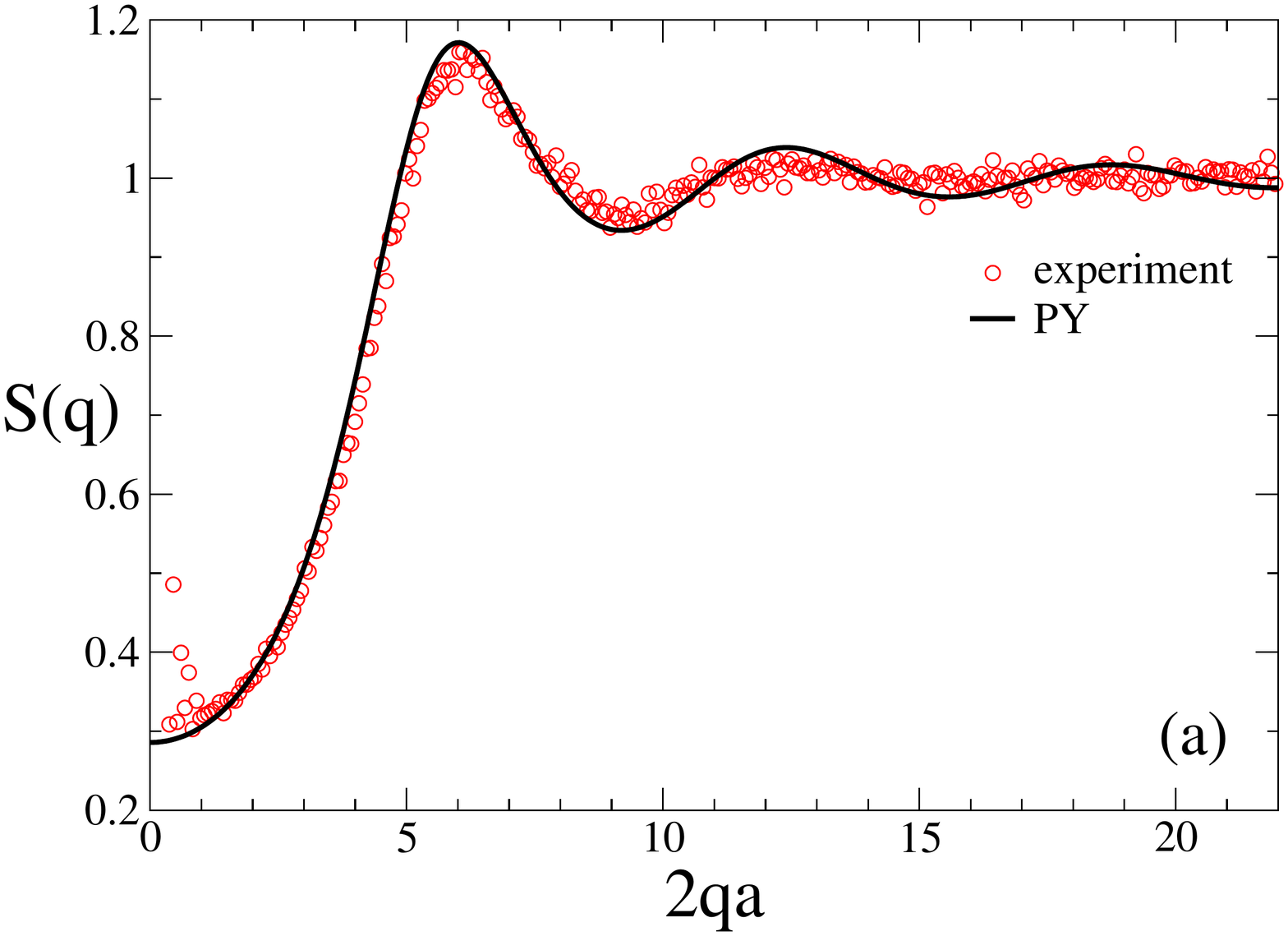}
\includegraphics[width=7cm,angle=0.,clip]{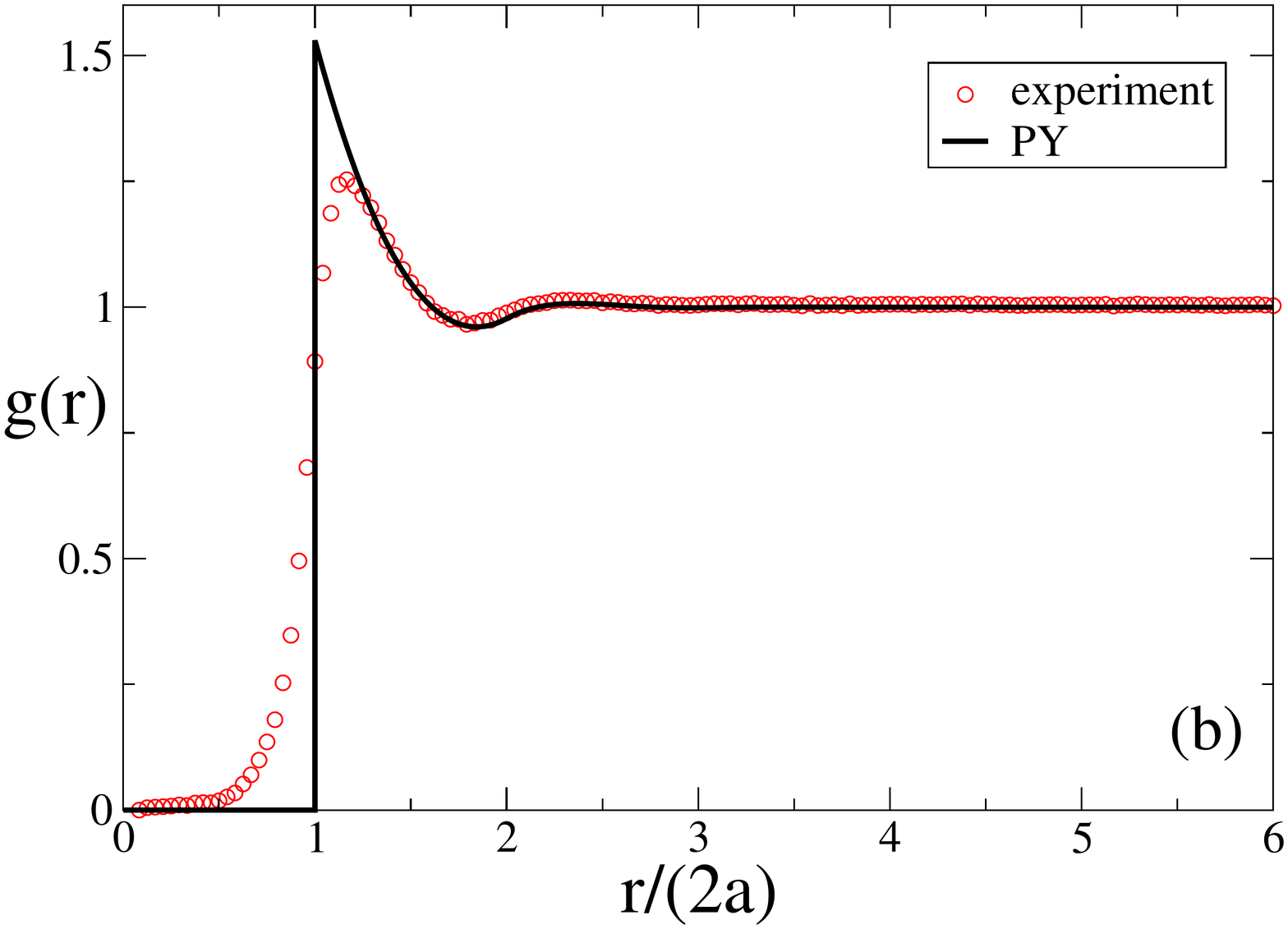}
\end{center}
\caption{ For $\phi=0.16$ and $c_p=0$, (a) static structure factor and
  (b) direct correlation function, calculated using the confocal
  microscopy coordinates (circles), compared to the PY calculations
  for hard spheres at the same $\phi$ (full lines). }
\label{fig:HS}
\end{figure}
This precise way to determine the static structure factor allows a
very stringent test on how close our colloids are to being a model
hard-sphere system.  In Fig.\ref{fig:HS}(a), we plot $S(q)$ for an
experimental sample with $\phi=0.16$ and $c_p=0$, together with PY
theoretical predictions at the same $\phi$.  The comparison also
permits an estimate of the particle radius which best reproduces the
superposition of the peak positions, which in this case turns out to
be $\simeq 600nm$, quite close to the one measured with dynamic light
scattering.  We observe a remarkable agreement of $S(q)$ with the
theoretical predictions essentially over all length-scales, which
demonstrates that, in the absence of polymer, our colloids behave as
hard spheres.  Similar results are found for the other investigated
$\phi$ values.  We also examine the direct correlation function
$g(r)$, the Fourier transform of $S(q)$\cite{hansen06}. Experimental
data and the PY prediction are shown for the same sample in Fig
\ref{fig:HS}(b). Because these fluid particles diffuse while 
being imaged, their motion provides a fundamental physical limit on how precisely 
their centers can be located. Our uncertainty from the image-processing software is
well under ± 50 nm in all directions\cite{Gao08a}. This positional uncertainty smears out the shape of $g(r)$, 
especially close to the contact peak, resulting in an apparent
worse agreement with the PY solution. We therefore conclude that the
$S(q)$, whose signal is maximum around $qa \approx \pi$, is a better
observable for direct quantitative comparison between confocal images
and theory/simulations.

\section{Locating the gel transition onto the thermodynamic phase diagram}
To locate the gel transition on the thermodynamic phase diagram, we
have employed a method based on the comparison of
experimentally-determined cluster mass distributions with those
obtained from simulations in the fluid phase up to gel
point\cite{Lu08a}. For each experimental sample (with a well-defined
$c_p$), its cluster mass distribution was determined, after defining
as connected any pair of particle whose distance is less than $1.3
\mu$m. The choice of this value, which is larger than the range of attraction, 
allows us to minimize spurious effects introduced by the aforementioned particle position 
uncertainties. Simulations were run at the same value of $\phi$ and
several values of $B_2^*$; in order to identify the optimal $B_2^*$
value, the cluster mass distribution that best matches 
the experimental data is identified via a least-squares minimization. Connected particles 
were defined with the same procedure in both experiments and simulations.
There are no free parameters
involved in this matching procedure, which uniquely associates each
experimental $c_p$ with a theoretical value of $B_2^*$. 
Because the experimental conditions are in the short-ranged
attractive limit, we can in principle use any short-ranged
attractive potential in the simulations. By using several
potential shapes (AO, square well and generalized
Lennard-Jones)---all of which yield quantitatively identical
results\cite{Lu08a}---we provide stringent reconfirmation of
the validity of the Noro-Frenkel law.
\begin{figure}[tbh]
\begin{center}
\includegraphics[width=10cm,angle=0.,clip]{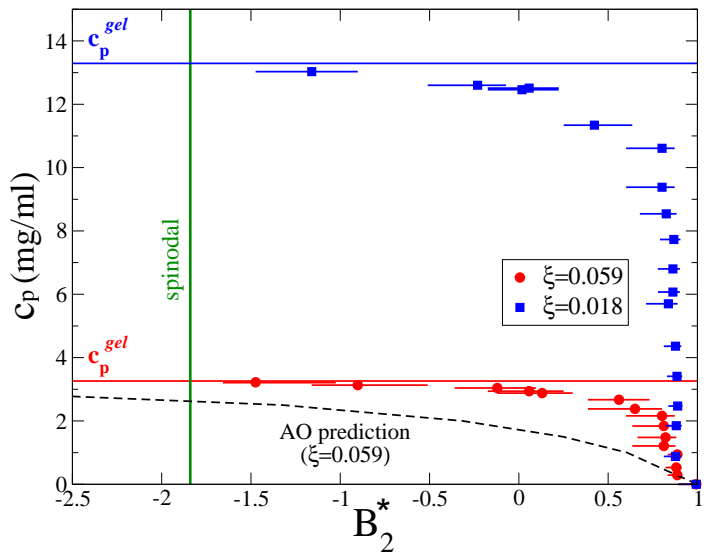}
\end{center}
\caption{Dependence of polymer concentration on the thermodynamic
  universal parameter $B_2^*$ for $\phi=0.045$ and two sets of
  polymers, with $\xi=0.018$ and $\xi=0.059$. The experimental gel
  transition is observed to happen exactly at the onset of
  thermodynamic phase separation.  The dashed line is the AO prediction for $\xi=0.059$.}
\label{fig:cp-B2}
\end{figure}

The results of this $c_p$-$B_2^*$ mapping procedure are presented in
Fig.~\ref{fig:cp-B2} for $\phi=0.045$ and the two studied sets of
polymers described above. Horizontal lines mark the
experimentally-observed concentration threshold $c_p^{gel}$ denoting
the transition from fluid to gel for each polymer. The vertical line
denotes the boundary of spinodal decomposition expressed as a value of
$B_2^*$, which depends only on $\phi$. This boundary is estimated both
within simulations, through the monitoring of the time-dependence of
the energy and of the small angle static structure
factor\cite{Zaccapri}, and via the PY energy route \cite{hansen06},
which has been shown to be quite accurate for predicting the spinodal
temperature\cite{Mil03a}; both values coincide within the error
bars. Without exception, the data reported in Fig. \ref{fig:cp-B2}, as
well as those for other studied $\phi$ (not shown here)\cite{Lu08a},
demonstrate that the gel transition takes place right at the
onset of spinodal decomposition.
(Note that the $B_2^*$ scale is
rather expanded when compared to a temperature scale, so that the
observed distance from the last fluid sample to the gel boundary is
very small, and comparable to the experimental error in $c_p$.)
These data suggest a mechanism for gelation of short-ranged attractive
particles: the system first enters the unstable region, where density
fluctuations drive the formation of locally-dense liquid droplets;
then these denser regions arrest probably through an intervening attractive
glass transition.  

Fig. ~\ref{fig:cp-B2} provides the `true' relationship between polymer
concentration and thermodynamic parameters: given the correct
potential model to describe the system, its $B_2^*$ should superimpose
onto these points. However, since $B_2^*$ is an integrated quantity of
the potential in terms of particle distance, the original potential
cannot be recovered easily from $B_2^*$ alone. However, if the simple
AO formula (Eq.\ref{eq:AO}) is used to determine $B_2^*$ as a function
of $c_p$ (dashed lines in Fig. \ref{fig:cp-B2}), the resulting
relation is rather different from what the cluster mass distribution
mapping predicts. To reach any given $B_2^*$, the AO model predicts a
much smaller value of $c_p$ than that observed in experiment,
demonstrating that it does not quantitatively describe the potential
for our experimental systems. The discrepancy, as expected, becomes
larger and larger for increasing polymer concentrations.

\section{Characterization of the Gel States}

We now examine some features of the observed gel states. For
each studied $\phi$, we have monitored the evolution over time of
several gel states at different $c_p$. In Ref.\cite{Lu08a} we have
reported $S(q)$, obtained from the particle coordinates, across the
fluid-gel transition. We observed that while the fluid, even at the
highest $c_p$, retains a low value of $S(q \rightarrow 0)$, a big jump
of about two orders of magnitude in the low-$q$ peak of $S(q)$ is
found as soon as the system crosses to a gel state. Moreover, we
followed the evolution in time of this spinodal peak, finding an
initial slow coarsening process (compatible with a power law
$t^{-1/6}$), later interrupted by arrest, i.e. at long times the
structure of the system reaches a final state and remains subsequently
unchanged\cite{Lu08a}. These results are in good agreement with
numerical simulation studies in the arrested spinodal
region\cite{Zaccapri,Fof05b}.

\begin{figure}[tbh]
\begin{center}
\includegraphics[width=7.75cm,angle=0.,clip]{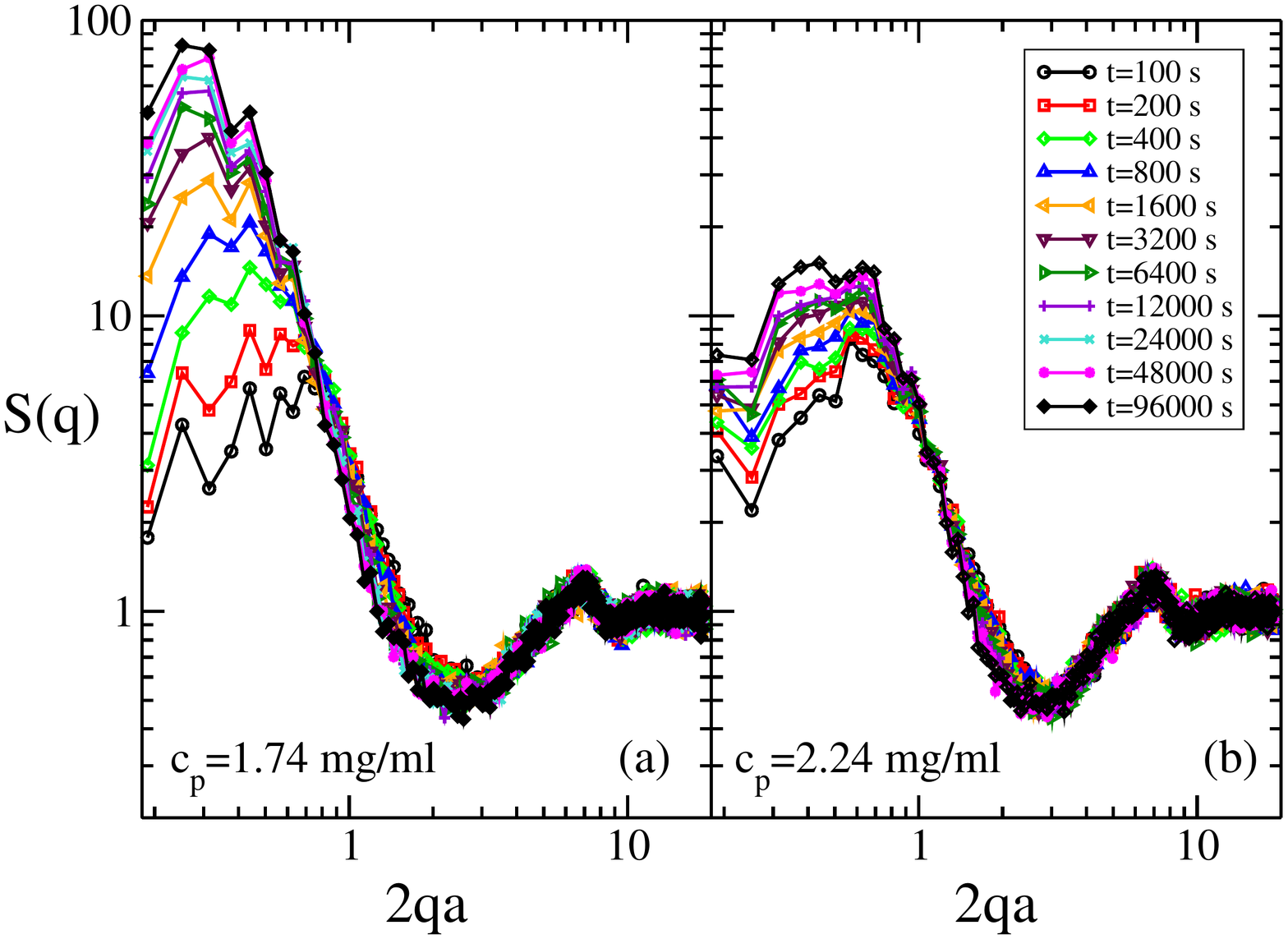}
\includegraphics[width=7.75cm,angle=0.,clip]{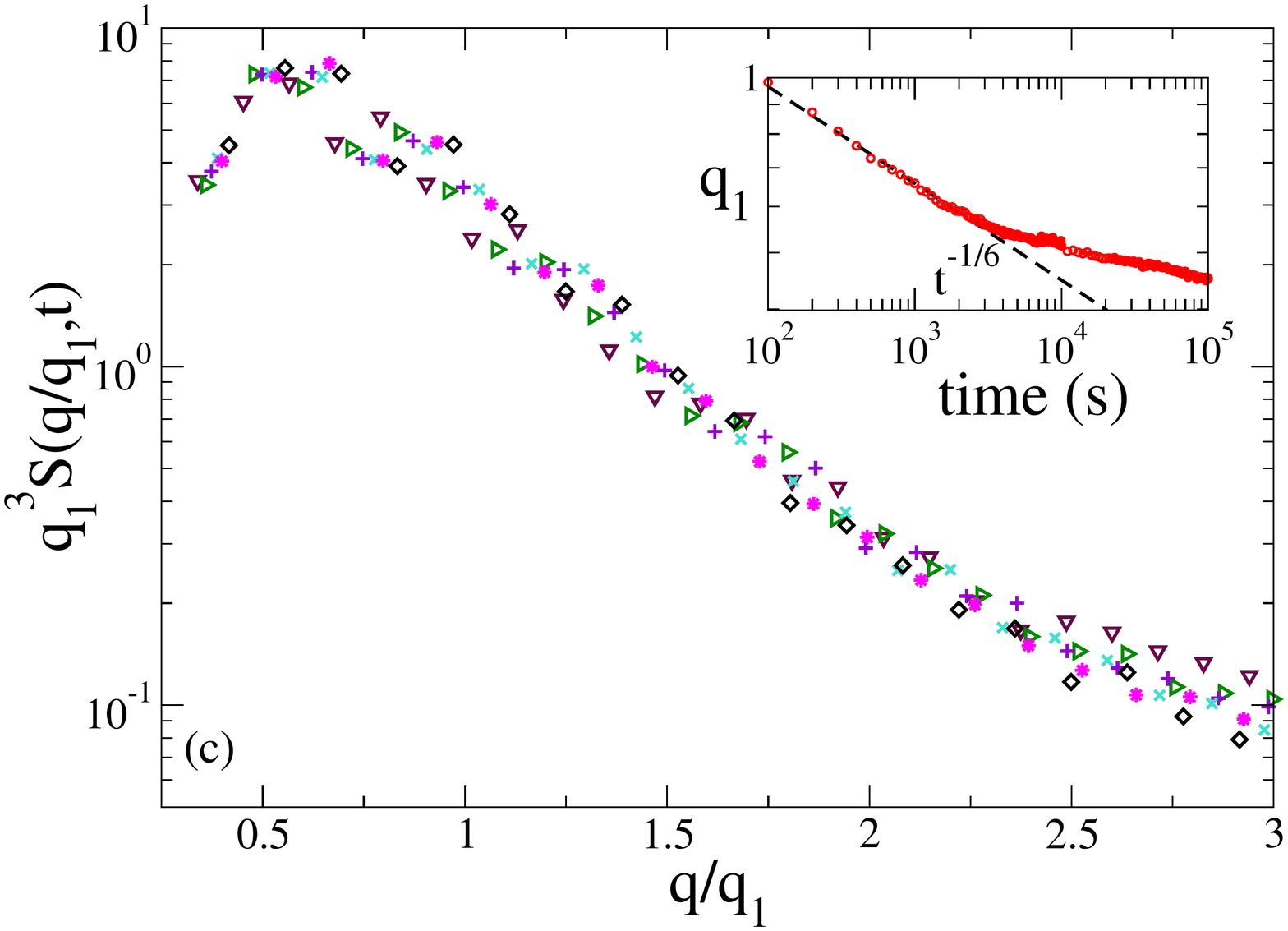}
\end{center}
\caption{ (a) Shallow versus (b) deep quench evolution in time of $S(q)$ for $\phi=0.16$. 
(c) Scaling of $S(q)$ as in 
  \protect\cite{Car92a} for the shallow quench case reported in (a) and $t\gtrsim 3000 s$. 
  The inset shows the time dependence of $q_1$ for the same state point; at short times the reduction of the characteristic length scale is compatible with $t^{-1/6}$, later followed by a slower decrease.
  Symbols are consistent throughout all panels. }
\label{fig:sq-phi016}
\end{figure}

We here complement these results with the additional evidence that the
arrested structure crucially depends on the location of the studied
state point within the two-phase region. Indeed, we plot in
Fig.\ref{fig:sq-phi016}(a) and (b) the evolution of $S(q)$ with time for
$\phi=0.16$ and two different $c_p$ values. On the left panel, the
data correspond to a system which is just inside the gel boundary,
i.e. a (shallow) quench just within the spinodal region. In this case, the
low-$q$ peak of $S(q)$ undergoes quite a significant coarsening
process, growing in amplitude by almost two orders of magnitude,
before eventually arrest takes place and the growth stops at long times. On
the right panel, we report data corresponding to a much deeper
quench inside the spinodal region. Here, the coarsening process almost
does not proceed at all, and the structure stops changing quite
rapidly, without even developing a large compressibility. Also the
peak position does not move significantly with time. These results
show that the resulting gels formed by arrested spinodal can differ very much in
their final structure, and in particular in their mesh size and compressibility, depending on the
chosen quench path inside the two-phase region.

We also monitor the evolution of $S(q)$ using the
scaling form for spinodal-driven aggregation derived by Furukawa\cite{Fur85a}, 
\begin{equation}
S(q/q_1, t)\sim q_1(t)^{-d} F(q/q_1),
\label{eq:furu}
\end{equation}
which was also generalized to the case of fractal clusters undergoing
DLCA by Carpineti and Giglio\cite{Car92a}.  Here as typical length
scale we use $q_1(t) \equiv \left(\int_0^{q_c} S(q,t) q dq\right)/\left(\int_0^{q_c} S(q,t) dq\right)$ where $q_c=6a$ is an appropriate cut-off value, 
which provides the first moment of
the wave-vector in the low-$q$ peak of $S(q)$ to reduce numerical
noise, while $F(q/q_1)$ is a time-independent scaling function and $d$
is the spatial dimension.  For fractal aggregation this is found to be
replaced by the fractal dimension of the clusters\cite{Car92a}.
Within our time window of observation of the gel states, three
different time intervals can be identified: an early stage (which we
observe only partially, as our first observation time is equal to
$100s$), an intermediate regime where the Furukawa scaling should hold; 
and, finally, the late stage or arrested regime, where the system
structure does not evolve any longer. In order to visualize the
intermediate regime and to demonstrate the scaling of Eq.\ref{eq:furu} with $d=3$,
we focus on state points not too far from the spinodal
boundary, where we observe a significant evolution of $S(q)$ over time.  
We find that our data, under these shallow quench
conditions, do scale onto a single master curve at intermediate times,
well before arrest takes place, as predicted by the
scaling. Trivially, data from later times also scale onto the same
master curve.  As an example we show in Fig.\ref{fig:sq-phi016}(c)
the scaled data corresponding to the shallow quench of
Fig.\ref{fig:sq-phi016}(a). For this particular state point, a
significant evolution of $S(q)$ over time is observed, as it is
evident from the time-dependence of $q_1$ reported in the inset. For
$t\gtrsim 3000 s$, which interestingly appears to be very close to
the time where $q_1$ departs from the $t^{-1/6}$ law towards a
slower form (until saturation, not observed just for this particular
sample), the scaling holds at all subsequent times. For all other
studied state points, arrest takes place much earlier so that the
scaling can be demonstrated only in a narrower regime. In all cases, the
scaling does not providence evidence of a fractal structure of the
clusters.

\begin{figure}[tbh]
\begin{center}
\includegraphics[width=7.75cm,angle=0.,clip]{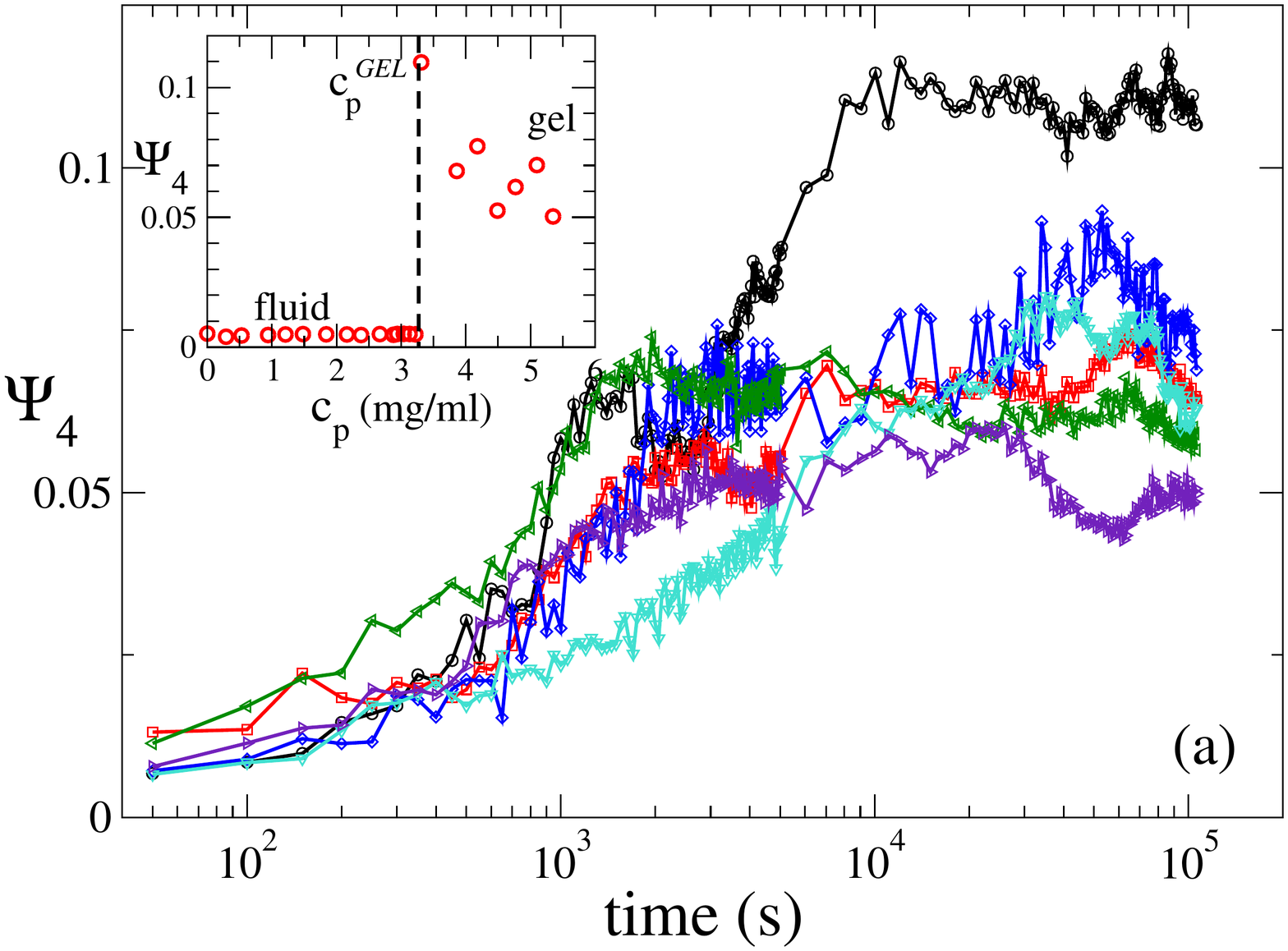}
\includegraphics[width=7.75cm,angle=0.,clip]{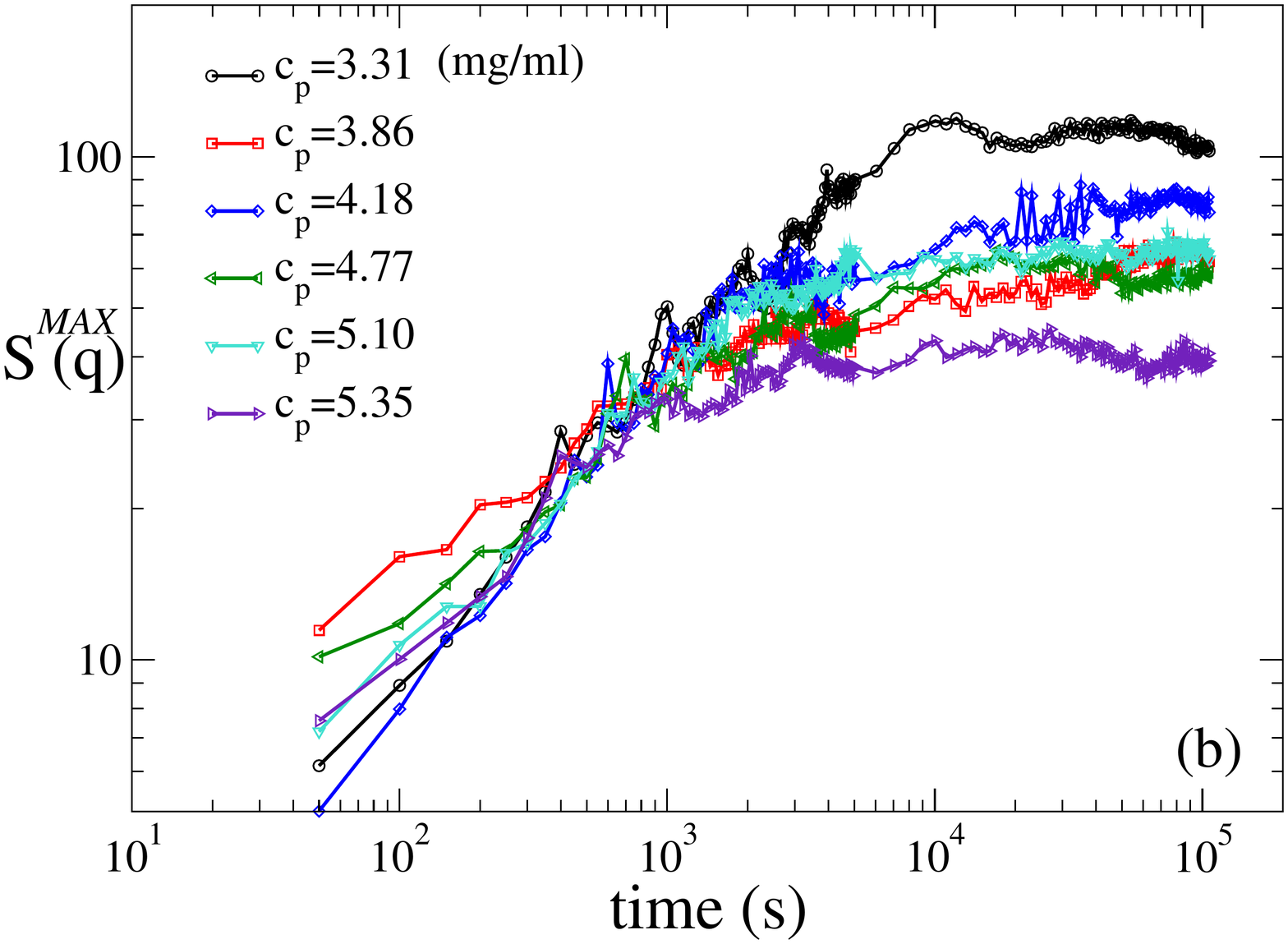}
\end{center}
\caption{(a): Demixing parameter $\Psi_4$ evolution with time for the
  gel states with $\phi=0.045$. Inset: the fluid to gel transition in
  terms of $\Psi_4$; (b)$S^{MAX}(q)$, the maximum height of $S(q)$,
  evolution with time for the same state points.  Identical lines and
  symbols are used in both panels. }
\label{fig:phi4}
\end{figure}

Another measure of the inhomogeneous character of the arrested states
is the so-called demixing (or inhomogeneous) parameter
$\Psi_n$\cite{Pue03a}, which can be evaluated by dividing the system
into $n^3$ boxes and measuring the local density in each box.  Hence,
\begin{equation}
\Psi_n\equiv\sum_{k=0}^{n^3} |\rho_{k}-\rho|^2,
\label{eq:psi4}
\end{equation}
 where $\rho_k$ is the density in the $k$-th box and $\rho=3\phi/(4\pi
 a^3)$ is the average particle number density. In a fluid or
 homogeneous state, this parameter averages to zero, while as soon as
 inhomogeneities appear, it grows. A comparison of $\Psi_n$ for
 different attraction strengths can provide information on how the
 degree of inhomogeneity varies with different quench paths.  
 We show
 results for $\Psi_4$---calculated by dividing the imaged volume of
 $(52\mu m)^3$ into $4^3 = 64$ boxes of $13\mu$m edge length---versus time, in
 Fig.\ref{fig:phi4}(a) for $\phi=0.045$ and all studied quench depths.
 We also repeated our analysis for a smaller box edge (of $4\mu$m),
 finding the same qualitative results.   The evolution in time
 of $\Psi_4$ for all gel samples displays an initial growth followed
 by a plateau region, similar to what observed for the maximum height
 of the low-$q$ peak of $S(q)$, named $S^{MAX}(q)$, for the same state
 points, reported in Fig.\ref{fig:phi4}(b).
In the inset of Fig.\ref{fig:phi4}(a) the average (for fluid states)
and long-time limiting value (for gel states) of $\Psi_4$ is reported, in order
to show, once again, the crossing of the spinodal region in
correspondence of the gel transition. It is intructive to notice that,
apart from the gel state very close to the spinodal boundary
(i.e. with $c_p=3.31 mg/ml$), whose spinodal decomposition process
seems to proceed further than the others before finally arresting, all
other gel states, corresponding to deeper quenches, seem to converge
toward a similar value of the inhomogeneity parameter.  Similar
results are also found for $\phi=0.16$.

\begin{figure}[tbh]
\begin{center}
\includegraphics[width=7.75cm,angle=0.,clip]{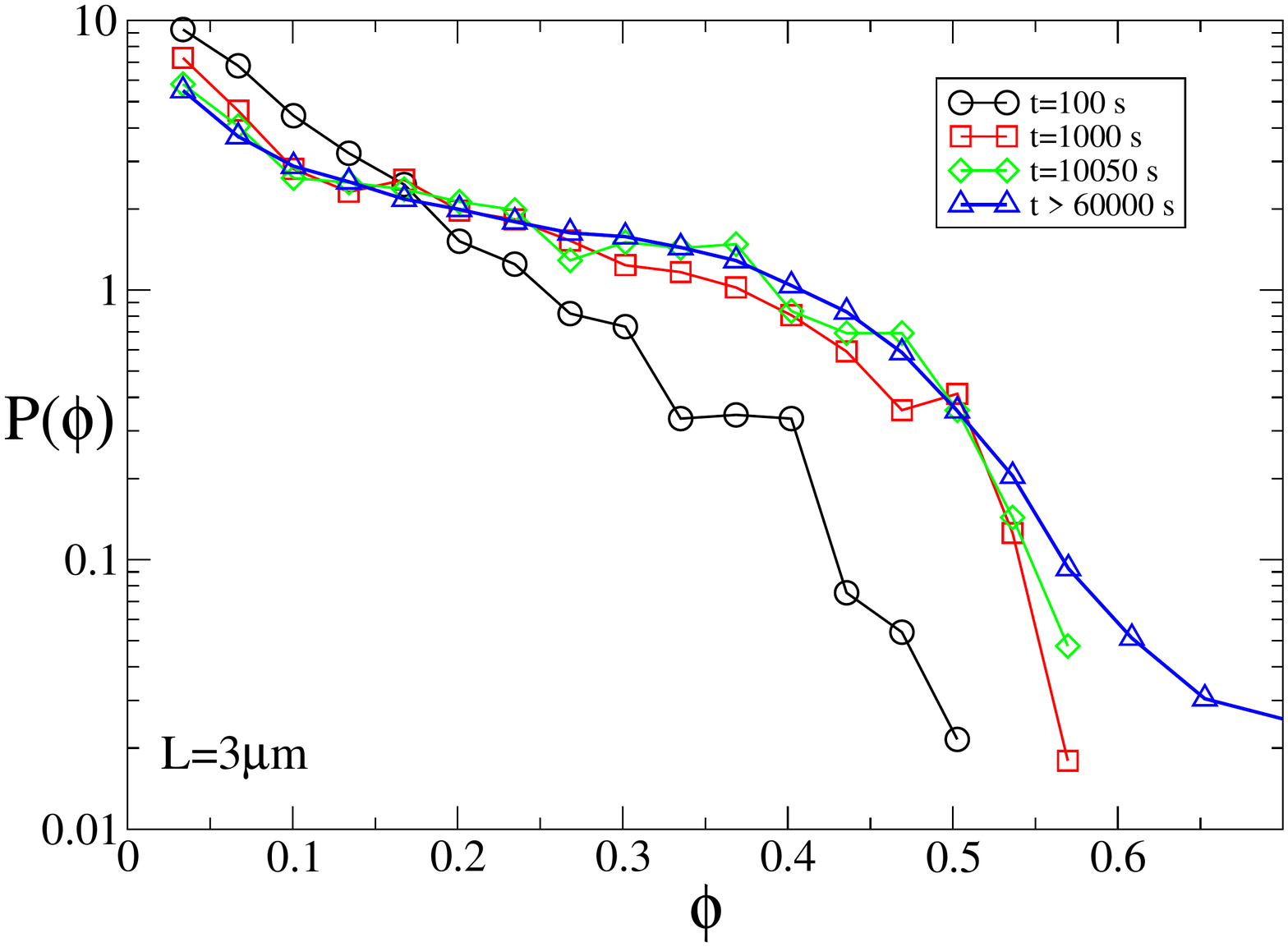}
\includegraphics[width=7.75cm,angle=0.,clip]{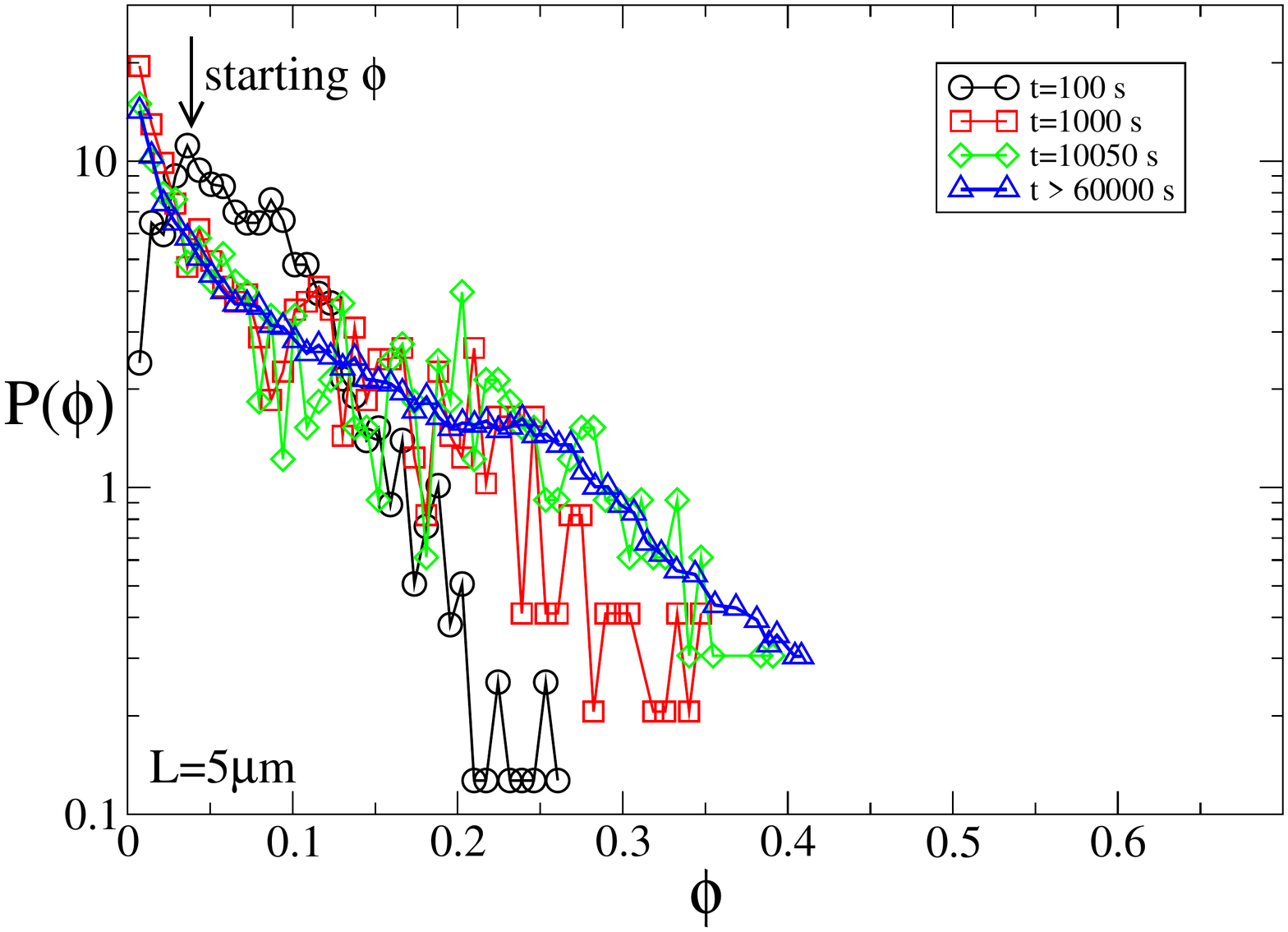}
\end{center}
\caption{Evolution in time of local packing fraction probability $P(\phi)$,
  obtained measuring the local $\phi$ in small boxes of sizes $3$ (left) and 5 (right)
  $\mu$m), for the
  state point $\phi=0.045 $ and $c_p=3.31$ mg/ml. }
\label{fig:prho}
\end{figure}

Finally, we examine the local density (or $\phi$) distribution
for several values of the box size. This
approach is sometimes implemented in simulations to establish the
presence of a two-phase region\cite{Bab06a}: if two distinct peaks
appear, then two phases of different densities can be properly
identified. The observation of the two peaks is strongly connected to
the choice of a box size comparable to the size of the spatial density
fluctuations, and requires a significant evolution of the phase
separation process to be observed\cite{Bab06a}. Fig. \ref{fig:prho}
shows the evolution in time of $P(\phi)$ for two different box sizes,
respectively $L=3 \mu m$ and $L=5 \mu m$ for $\phi=0.045$
(corresponding to an average number of particle per box of
respectively 1.5 and 7.1).  Results show that the system starts from a
rather homogeneous condition (since $P(\phi)$ on the $L=5 \mu$m scale
shows a maximum around the average $\phi$), then evolves 
in the direction of increasing the
density fluctuations (since states with both small and large packing
become preferentially probed) until arrest is reached.  On the small
$L=3$ scale, the local density increases even more than a factor 15,
reaching values comparable to a dense packing of spheres ($\phi >
0.6$), consistent with the estimates of the internal gel density
derived in Ref.~\cite{Lu08a}.  Smaller $\phi$ values are probed on the
$L=5$ scale, suggesting that the characteristic mesh of the aggregate
is comparable to the smaller box size.  We also note that, in the time
window preceding dynamic arrest, we do not observe the formation of a
bimodal distribution, suggesting that gel formation may precede true bulk phase separation.

\section{Conclusions}

In this manuscript, we have provided additional evidence of the
mechanism at the origin of the gel transition in short-ranged
attractive colloid-polymer mixtures.  We have firstly shown that
the experimental colloidal
system is well characterized: without added polymers, it
behaves as a system of nearly-perfect hard spheres. Then, comparing
cluster distributions in experiments and simulations has allowed us
to precisely map the experimental gel transition on the
thermodynamic phase diagram, quantifying the attraction
strength in terms of the normalized second virial coefficient $B_2^*$.
This universal parameter contains all the necessary information
for short-ranged isotropic attractive potentials to entirely
determine the structure and connectivity properties of the
system. In this way, our procedure is independent of any
specific choice of interaction potential model and is relevant
to any short-ranged isotropic attractive system. Our findings
unambiguously show that thermodynamic phase separation triggers
the kinetic arrest\cite{Lu08a}.

A growing consensus has recently favored the arrested phase
separation mechanism for gelation in these short-ranged
attractive systems. A number of simulations
studies\cite{Zaccapri,Fof05b,Cha07a} and various experimental
colloid investigations---including colloids with depletion
interactions\cite{Man05a,Buz07a} as well as mixtures of oppositely
charged colloids\cite{San08a}---all support this picture, which
may also be relevant for atomic glass-formers\cite{Sas00PRL}.
Moreover, our results confirm important conclusions drawn from
data on attractive lysozyme solutions, recently reported by
Cardinaux et al.\cite{Car07a}. Both in our colloidal system and
in the protein solution case, gelation is caused by the arrest
of the dense phase induced by the liquid-gas spinodal
decomposition. Yet differences between the two systems appear
to arise in the estimated value of the gel packing fraction;
this packing fraction appears to be quench-depth independent as
evaluated in Ref.~\cite{Lu08a} and further confirmed by the present analysis, 
while dramatically depending on
quench depth in lysozyme, where the density of the dense phase
shows a marked decrease with increasing attraction
strength\cite{Car07a}. However, several factors may contribute to these
reported differences. First, differences in the interaction
potential are present between lysozyme and colloid-polymer mixtures.
In lysozyme, electrostatic repulsion\cite{Card06} can play an important role
and, in principle, anisotropy effects (due to shape or
patchyness) could be relevant. Indeed, one observes a rather
low-$\phi$ position of the critical point ($\phi_c \approx
0.18$\cite{Car07a}) as compared to standard short-ranged
attractive potentials ($\phi_c \simeq 0.27$\cite{Lar08a}).
Second, the methods used to establish the local gel volume
fraction are rather different in the two cases. While we have
used a microscopic method, based on single-particle location, to estimate
the internal volume fraction of the gel \cite{Lu08a}, Cardinaux
et al. determine the gel volume fraction by centrifuging the
lysozyme solution after the phase separation was arrested,
decanting the supernatant, and measuring the resulting macroscopic volume fraction.
Whether these two metrics would yield the same results, if they
could both be performed on the same sample, it is presently
unknown. A further complication arises by the fact that single
$3.4$ nm lysozyme proteins cannot be individually resolved.
Finally, the finite
size of the imaged volume in our experiment, as compared to the
macroscopic volume (compared to particle size) accessed in
\cite{Car07a}, could also be used as an explanation for the
discrepancy, although we performed additional checks for larger samples of 
millimetric size confirming our previous findings.

Another interesting question relates to why we do not observe
any state point undergoing a complete gas-liquid phase
separation within our study. In all experimentally studied
cases for samples that begin to phase separate, we find that a
gel forms at long times. Indeed, for such short-ranged
attractive systems at the values of $\phi$ that we study, the
difference in attraction strength that would allow to explore
the liquid region is very small and roughly comparable to the
experimental error in $c_p$. However, we do observe some state
points, very close to the spinodal boundary where arrest only
arises at very long times (see Fig.\ref{fig:sq-phi016}).
Also, it appears that our studied state points always refer to quenches inside the spinodal, where
the liquid droplets rapidly coalesce and form an arrested spanning network. Another interesting case would be to look at quenches
inside the region at low $\phi$ delimited between binodal and spinodal lines, where unconnected clusters may
become arrested, forming the so-called glassy `beads'\cite{Cat04aJPCM,Sedg05}.

Finally, we discussed in some detail the properties of the
observed gel states. We have provided evidence, from complementary
types of analysis, that the resulting states are rather
inhomogeneous, that their final structure depends on the quench
path, and that locally the internal gel volume fraction is very
high \cite{Lu08a}. Kinetic arrest intervenes before phase
separation has proceeded enough to allow the visualization of
two distinct peaks in the local density
distribution\cite{Bab06a}, and the system becomes trapped in a
highly heterogeneous state. Future work will be devoted to
attempts to validate our method at larger packing fractions and
to test its limit of validity with increasing range of
attraction.

\section{Acknowledgments}
EZ and PJL equally contributed to the present work.  We thank
A. B. Schofield for the synthesis of the PMMA particles and
P. Schurtenberger and E. Sanz for useful comments. We acknowledge
support from Marie Curie Network on Dynamical Arrest of Soft Matter
and Colloids MRTN-CT-2003-504712, the NoE SoftComp
NMP3-CT-2004-502235, NASA (NAG3-2284), the NSF (DMR-0602684) and the
Harvard MRSEC (DMR-0820484).

\section{References}
\bibliographystyle{./iopart-num} 
\bibliography{./articoli,./altra,./advances,./biblio_patchy.bib,./star-star}

\end{document}